\documentstyle[epsfig]{aipproc}

\def\msun{$M_{\odot}$}

\def\etal{{\it et al.}}

\def\asec{\ifmmode ^{\prime\prime}\else$^{\prime\prime}$\fi}
\def\amin{\ifmmode ^{\prime}\else$^{\prime}$\fi}
\def\degs{\ifmmode ^{\circ}\else$^{\circ}$\fi}
\newbox\grsign \setbox\grsign=\hbox{$>$}
\newdimen\grdimen \grdimen=\ht\grsign
\newbox\laxbox \newbox\gaxbox
\setbox\gaxbox=\hbox{\raise.5ex\hbox{$>$}\llap
     {\lower.5ex\hbox{$\sim$}}}\ht1=\grdimen\dp1=0pt
\setbox\laxbox=\hbox{\raise.5ex\hbox{$<$}\llap
     {\lower.5ex\hbox{$\sim$}}}\ht2=\grdimen\dp2=0pt

\def\msun{$M_{\odot}$}

\begin{document}

\title{RXTE Observations of GRS 1915+105}

\author{J. Greiner$^*$, E.H. Morgan$^{**}$, R.A. Remillard$^{**}$}
\address{$^*$Astrophysical Institute Potsdam, 14482 Potsdam, Germany \\
         $^{**}$Center for Space Research, MIT, Cambridge, MA 02139, USA}

%\lefthead{LEFT head}
%\righthead{RIGHT head}
\maketitle

\begin{abstract}
We report on extensive X-ray observations of 
the galactic superluminal motion source GRS 1915+105 
with the RXTE satellite over the last year.
More than 130 RXTE pointings have been performed on roughly a weekly basis.

GRS 1915+105 displays
 drastic X-ray intensity variations on a variety of time scales
ranging from sub-seconds to days. 
In general, the intensity changes are accompanied by spectral changes on the
same timescale. Three types of  bursts with typical durations
between 10--100 sec have been identified
which have drastically different spectral properties and seem to occur in
a fixed sequence.
One of the most intense bursts has a bolometric X-ray luminosity
of $\approx$5$\times$10$^{39}$ erg/s during the 25 sec maximum-intensity part.

\end{abstract}

\section*{Introduction}

GRS 1915+105 was discovered on 1992 August 15  with the WATCH detectors 
on {\it Granat} (Castro-Tirado \etal\ 1992).
A comparison of the BATSE ($>25$ keV) flux with that of ROSAT (1--2.4 keV)
fluxes obtained during regularly performed pointings has shown that 
GRS 1915+105 has been active all the time, even during times of BATSE
non-detections (Greiner \etal\ 1997).
A variable radio source was found with the VLA (Mirabel \etal\ 1993a)
inside the $\pm$10\asec\ X-ray error circle (Greiner 1993), which later was 
discovered to exhibit radio structures travelling at apparently 
superluminal speed (Mirabel \& Rodriguez 1994) making GRS 1915+105 the first
superluminal source in the Galaxy. Until then, apparent superluminal motion
was only observed in AGN, the central engines
of which are generally believed to be massive black holes. This similarity
suggests that GRS 1915+105 harbors a stellar-sized black hole.

The X-ray spectrum as seen with  ROSAT (Greiner 1993) and ASCA (Nagase \etal\
1994) is strongly absorbed (N$_{\rm H}\approx 5\times10^{22}$ cm$^{-2}$) 
consistent with the location in the galactic plane at 12.5 kpc distance
(Mirabel \& Rodriguez 1994).

\section*{Pointed RXTE Observations}

The first two RXTE observations were performed on April 6 and 9, 1996.
The surprising results of these observations and those of the daily RXTE
ASM dwells triggered an unique sequence of RXTE pointings on
GRS 1915+105 on a roughly weekly time scale. These data are publicly available
and a number of papers have already appeared dealing with various aspects
of the extremely rich variety of X-ray properties of GRS 1915+105:
(Greiner, Morgan and Remillard 1996; Chen, Swank and Taam 1997; Belloni \etal\ 
1997; Morgan, Remillard and Greiner 1997; Taam, Chen and Swank 1997).
\vspace{-0.1cm}

\subsection*{Temporal characteristics}

The X-ray light curves reveal
a variety of features, one of which is large amplitude intensity variations.
We identify the following properties in the light curves of GRS 1915+105:
\begin{itemize}
\vspace{-0.28cm}\item 
%Repeating pattern of sputters: 
In 15 of the pointed PCA observations, we find large, eclipse-like dips in 
the X-ray flux, which we call sputters. During these sputters the flux 
drops from $\approx$2--3 Crab to a momentary lull at 
about 100--500  mCrab and
then shoots up again. The spectrum softens dramatically during the sputters,
thus arguing against absorption effects. 
\vspace{-0.28cm}\item 
On some occasions we see extremely large amplitude oscillations with an
amplitude of nearly 3 Crab and periods of 30--100s.
\vspace{-0.28cm}\item Between the episodes of large-amplitude variations
the X-ray flux variations are more regular, developing into
clearly visible quasi-periodic oscillations which seem to be stable
over several days.
\vspace{-0.28cm}\item The combination of the intense
QPOs and the high throughput of the PCA enabled phase tracking of
individual oscillations: the QPO arrival phase
(relative to the mean frequency) exhibits a random walk with no
correlation between the amplitude and the time between subsequent
events.  Furthermore, the mean `QPO-folded' profiles are roughly
sinusoidal with increased amplitude at higher energy, and with a
distinct phase lag of $\approx$0.03 between 3 and 15 keV. 
\vspace{-0.25cm}
\end{itemize}

\subsection*{Spectral characteristics}

\begin{figure}[t!] 
\centerline{\epsfig{file=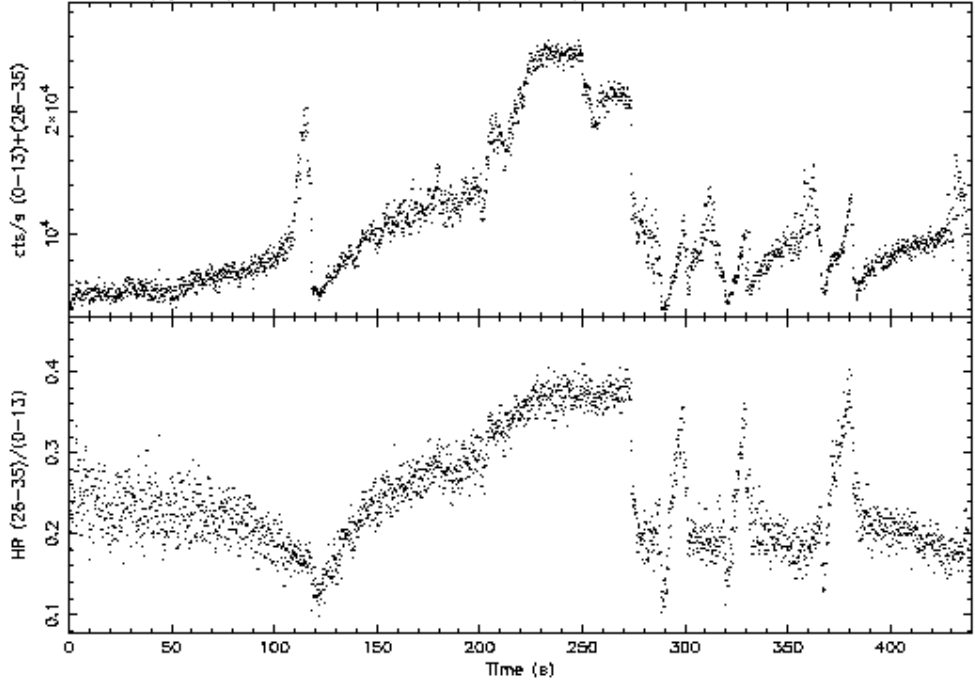,width=0.95\textwidth}}
%   bbllx=1.8cm,bblly=2.3cm,bburx=17.cm,bbury=27.3cm,clip=}}
\caption{Part of the lightcurve as observed on October 13, 1996. The lower 
   panel shows the countrate (hardness) ratio of channels 26--35 
  (10.6--12.2 keV) versus
  channels 0--13 (3.1--5.3 keV) while the top panel shows the summed 
   countrate of these two 
   bands at 0.2 sec resolution (see text for more details). }
\label{flare}
\end{figure}

We have started a comprehensive spectral investigation using PCA as well
as HEXTE data of well-defined
time stretches which are selected according to their different shapes
in the lightcurve. While the work is still in progress we note the 
following, more general properties:

\begin{itemize}
\vspace{-0.28cm}\item The spectra are complex and rapidly variable. 
Single component spectra like pure power law,
bremsstrahlung, synchrotron or comptonisation models do not fit these spectra. 
In general, the spectra are composed of at least two components: one soft
component extending up to about 15--20 keV, and a flat, hard component
extending up to 200 keV which is well represented by a power law of
photon index 2.5--3.5. The hardness ratios 
demonstrate that the spectrum varies on timescales of seconds!

\vspace{-0.28cm}
\item The soft component can be well described by either an exponentially 
cut-off power law, a bremsstrahlung model, or disk blackbody models.
Given typical X-ray luminosities during most of the April to November 1996
activity state of 10$^{39}$ erg/s, and an upper limit for the size of the
emission region defined by the observed spectral changes on timescales of
seconds, the bremsstrahlung model is physically excluded
(emissivity is too low). We therefore use a multicolor disk blackbody spectrum
(DISK in XSPEC).

\vspace{-0.28cm}\item
The spectrum of the X-ray emission during the  lulls
is softer than during the high-intensity states, i.e. these lulls are 
not caused by absorption or any low-energy cut-off.
We have selected photons (for individual layers and single PCA units)
at different time intervals corresponding to these two intensity states.
The gross energy distribution of the high-intensity emission  can 
be described by a disk blackbody temperature of 1.9 keV and a power law 
with photon index $\alpha = -2.6$.
The spectrum during the lulls is well represented
by a disk blackbody with a temperature of 1.1 keV plus a power law of photon 
index $\alpha = -2.3$.

\vspace{-0.28cm}\item
As can be inferred from the hardness ratios, there are no major
spectral changes during the decay phase between lulls before the onset of
the large-amplitude oscillations. But during these oscillations the 
temperature oscillates on the same time scale as the intensity.
The power law model has a photon index of --2.6, and nothing can be said
on rapid variability of this component due to low statistics on these
1--2 sec timescales.

\vspace{-0.28cm}
\item We identify three burst-like events with completely different
spectral behaviour (see Fig. \ref{flare}). 
First, bursts 
with a typical disk temperature of $\approx$1.1 keV and little changes
along the burst (though a slight, smooth softening seems possible;
see Fig. \ref{flare} at t=110--120 sec). However,
at the end of these {\it type 1} bursts the spectrum changes abruptly and 
starts hardening.
Second, bursts with generally longer timescale and larger amplitude
(see lower panel of Fig. \ref{flare} at t=150--280). They show a smooth
hardening despite more erratic intensity variations on top of the general
flare profile.
Third, bursts with a temperature variation proportional to the intensity
(see lower panel of Fig. \ref{flare} at t=290--300, 325--335 or 370--380 sec)
and a maximum temperature of $\approx$2 keV ({\it three} bursts).
In general, we find the following sequence: one type 1 burst preceding a
major flare (type 2 burst), followed by a series of alternating type 1 
and type 3 bursts. We have identified at least 6 such sequences during
the October 13--25, 1996 period. It is interesting to note that the type 1
bursts are similar to those occuring during or at the end of prolonged lulls 
like those of May 26, 1996 (see Fig. 4 of Greiner \etal\ 1996) and April 6
(ibid. Fig. 3) or Oct. 7, 1996 (Fig. 1 of Belloni \etal\ 1997).

\vspace{-0.28cm}
\item The major flare (type 2 burst) on October 13, 1996 (Fig. \ref{flare})
 is the most intense
emission we have detected from GRS 1915+105 so far. The nearly flat-top
main peak has a duration of nearly 30 sec., and integrating over the 
2 keV disk blackbody plus the --2.6 power law up to 100 keV results
in an unabsorbed luminosity (at the adopted distance of 12.5 kpc)
of 5$\times$10$^{39}$ erg/s.

\end{itemize}

\section*{Discussion}

The lack of coherent pulsations, the strong
variability during the high-intensity states on time scales well below one 
second and the impossibility of a bremsstrahlung interpretation 
indicate that the emission originates in an accretion disk.

The drastic intensity variations were interpreted as an inherent accretion
instability, rather than absorption effects, since there was spectral
softening during these dips.  The repetitive, sharp variations
and their hierarchy of time scales are entirely unrelated to the
phenomenology of absorption dips (Greiner \etal\ 1996).
The nature of these astonishing X-ray instabilities is
currently a mystery though attempts have been made to both interprete these as
accretion disk instabilities leading to an infall of parts of the
inner accretion disk (Greiner \etal\ 1996, Belloni \etal\ 1997) and relate
them to radio flares (Greiner \etal\ 1996, Pooley \& Fender 1997).

At photon energies above 10 keV, the high amplitudes and sharp profiles of the
QPOs are inconsistent with any scenario in which the phase delay is
caused by scattering effects. Alternatively, it appears that the
origin of the hard X-ray spectrum itself (i.e. the creation of
energetic electrons in the inverse Compton model) is functioning in a
quasiperiodic manner. These results fundamentally link X-ray QPOs with
the most luminous component of the X-ray spectrum in GRS 1915+105.

Taam \etal\ (1997) have investigated PCA data of October 15, 1996 and
found that type 1 and type 3 bursts always occur together. Our finding confirms
this behaviour with the addition that such sequences are always preceded by
one pair of type 1 and 2 burst.

The maximum (during a flare on Oct. 13, 1996) unabsorbed X-ray luminosity 
is 5$\times$10$^{39}$ erg/s. This is well
above the Eddington luminosity for a neutron star with any reasonable  mass,
suggesting that the system contains  a black hole. Assuming that the
emission is near (but slightly below) the Eddington luminosity, the
inferred mass (35 \msun) is compatible with that derived from the 
stable 67 Hz QPO (Morgan \etal\ 1997).

\vspace{0.1cm}{\it Acknowledgements:\,}{\small
 JG is supported by the German Bundesministerium f\"ur Bildung,
Wissenschaft und Forschung (BMBW/DARA) under contract Nos. 50 QQ 9602 3
and is grateful to DFG for a substantial travel grant (KON 1088/1997 and
GR 1350/6-1).}
 
\end{document}